\documentclass[preprint, 12pt,  authoryear]{elsarticle}
\usepackage{graphicx}
\usepackage[margin=1in]{geometry}
\usepackage{setspace}
\usepackage{amsmath, amssymb}
\usepackage{booktabs, multirow} 
\usepackage[authoryear]{natbib}
\newif\ifblinded
\ifdefined\journalversion \blindedtrue \fi
\newcommand{\ivate}{\ifblinded\texttt{ivate}\else\href{https://github.com/lulocher/ivate}{\texttt{ivate}}\fi}
\newcommand{\appfigname}{Figure}
\newcommand{\apptabname}{Table}
\usepackage{tikz}
\usepackage{xurl}
\usepackage[hidelinks]{hyperref}
\usetikzlibrary{arrows.meta}


\def\LATE{\Psi_{\text{LATE}}} %
\def\ATE{\Psi_{\text{ATE}}} %

\def\notInd{\not\!\perp\!\!\!\perp}

\newcommand{\sensAT}{\theta_A^{(0)}}
\newcommand{\sensNT}{\theta_N^{(1)}}

\newcommand{\thresholdAT}{\tau_A^{(0)}}      %
\newcommand{\thresholdNT}{\tau_N^{(1)}}      %
\newcommand{\thresholdNTstar}{\tau_N^{(1)\ast}}  %
\newcommand{\thresholdNTstarhat}{\hat{\tau}_N^{(1)\ast}}

\begin{document}

\begin{frontmatter}
    \title{Informative inference on the average treatment effect with instrumental variables}
    
    \author[1]{Luca Locher}
    \author[2]{Thomas Richardson}
    \author[3]{Mats Stensrud\fnref{fn:funding}}
    \fntext[fn:funding]{Mats Stensrud was funded by a starting grant from the Swiss National Science Foundation.}
    
    \address[1]{Epidemiology, Biostatistics and Prevention Institute, University of Zurich, Switzerland}
    \address[2]{Department of Statistics, University of Washington, USA}
    \address[3]{Institute of Mathematics, École Polytechnique Fédérale de Lausanne, Switzerland}

    \begin{abstract}
    Policy decisions often depend on whether the average treatment effect (ATE) exceeds a specific cutoff. In instrumental variable (IV) settings, point identification of the ATE requires strong assumptions that are often difficult to justify. Therefore, many researchers shift their attention to identification of the local (L) ATE or target bounds of the ATE. Both strategies have well-known drawbacks: the substantive relevance of the LATE is frequently questioned, and the bounds are often too wide to be informative. We propose a simple sensitivity analysis that allows researchers to maintain focus on the ATE, and then discuss its connection to related bounding procedures. Based on these results, researchers can decide, in a principled manner, whether the sign of the ATE is inferable under assumptions they deem credible. The proposed strategies are easy to apply, and we illustrate their value with two real-data examples. The R package \ivate{} implements our procedures.
    \end{abstract}
\end{frontmatter}

\onehalfspacing
\section{Introduction}
In instrumental variable (IV) settings, point identification of the average treatment effect (ATE) requires strong assumptions that are often difficult to justify in practice. Many researchers therefore shift their attention to an alternative estimand: the local average treatment effect (LATE), also known as the complier average causal effect (CACE) \citep{angrist_identification_1996}. The practical relevance of the LATE has been widely questioned \citep{deaton_instruments_2010, heckman_comparing_2010, manski_lure_2020, swanson_challenging_2018}. Indeed, when the ATE is of primary interest, a point-identified LATE gives only a partial answer to the substantive research question.

To maintain focus on the ATE under the same or weaker assumptions, partial identification approaches that target bounds on the ATE have been proposed; see \cite{swanson_partial_2018} for an overview in the binary setting. While partial identification methods have received substantial attention in the methodological literature, applications remain rare: a recent survey found that in a sample of more than 300 IV studies published in leading journals across epidemiology, medicine, political science, and economics, none applied partial identification methods \citep{locher_interpretational_2025}. One potential reason is that without additional assumptions, the resulting bounds represent extreme scenarios, thus are typically wide, and hence often fail to resolve the research question.

This paper addresses this tension by proposing a sensitivity analysis that allows researchers to assess whether credible inference on the sign of the ATE is feasible in a given application. Specifically, we derive thresholds for the latent quantities preventing point identification, expressed in terms of easily interpretable comparisons. We then discuss how this relates to procedures that translate restrictions on these quantities directly into improved bounds on the ATE. These strategies enable researchers to assess, in a principled manner, whether the available data and credible assumptions are sufficient to resolve the decision problem. Finally, we illustrate their value in two case studies.

\section{Informative inference on the ATE under credible assumptions}
\subsection{Setup}
\label{sec:setup}
We use capital letters to denote random variables and lower-case letters to denote their possible realizations. Consider a population of individuals indexed by $i = 1, \dots, N$. For each individual $i$, we observe an instrumental variable $Z_i \in \{0,1\}$, a treatment variable $A_i \in \{0,1\}$, and an outcome variable $Y_i \in \mathbb{R}$. We use parentheses to denote potential outcome variables. For example, $Y_i(z, a)$ denotes the potential outcome under an intervention that sets $Z_i=z$ and $A_i=a$, and $A_i(z)$ denotes the potential treatment under an intervention that sets $Z_i = z$.

Suppose first that the standard IV assumptions of \cite{angrist_identification_1996} hold:
\begin{align}
    &\text{(Consistency)} \label{ass:1} \\ 
    &\quad \text{(a)} \quad \text{If } Z_i = z, \text{ then } A_i = A_i(z), \text{ for all }  z, \text{ with probability 1},\nonumber \\
    &\quad \text{(b)} \quad \text{If } Z_i = z \text{ and } A_i = a, \text{ then } Y_i = Y_i(z, a), \text{ for all }  z, a, \text{ with probability 1},\nonumber \\
    & \text{(Independence)} \quad Z_i \perp\!\!\!\perp \{A_i(z), Y_i(a)\} \text{ for all } z, a \in \{0,1\},  \label{ass:2} \\
    & \text{(Exclusion restriction)} \quad Y_i(z, a) = Y_i(a) \text{ for all }  z, a, \text{ with probability 1}, \label{ass:4}\\
    & \text{(Monotonicity)} \quad A_i(z) \geq A_i(z') \text{ for all } z \geq z', \text{ with probability 1}, \label{ass:5} \\
    & \text{(Relevance)} \quad Z_i \notInd A_i \label{ass:3}.
\end{align}

The consistency assumption \eqref{ass:1} ensures that the potential treatments $A_i(z)$ are well-defined, partitioning individuals into four latent types according to $(A_i(0), A_i(1))$: compliers, always-takers, never-takers, and defiers. Monotonicity~\eqref{ass:5} rules out defiers, leaving three types. 

Let $T_i \in \{C, A, N\}$ denote the latent type of individual $i$, and let
\[
\pi_t = \mathbb{P}(T_i = t)
\]
denote the population share of type $t$.

Define the ATE as
\[
\ATE = \mathbb{E}[Y_i(1) - Y_i(0)],
\]
and the LATE as
\[
\LATE = \mathbb{E}[Y_i(1) - Y_i(0) \mid T_i = \mathrm{C}],
\]
i.e., the ATE among compliers. Under \eqref{ass:1}--\eqref{ass:5}, relevance \eqref{ass:3} ensures $\pi_C > 0$, so that the LATE is well defined. However, relevance is not strictly required for the procedures introduced in this article.

Suppose that a decision maker uses the observed data to inform a policy decision based on the criterion
\begin{equation}
    \ATE > x,
    \label{eq:criterion}
\end{equation}
where $x \in \mathbb{R}$ is a known policy-relevant threshold. 
So that, for example, if the ATE exceeds this threshold, the treatment is deemed effective.
While under \eqref{ass:1}--\eqref{ass:3} the LATE is point identified, the ATE is not.

\subsection{Remark on cross-world assumptions}
\label{sec:remark_cw}
The monotonicity assumption which enables point identification of the LATE is cross-world, and thus not consistently falsifiable, since it depends on features of the joint distribution of potential outcomes that can never be observed \citep{richardson_single_2013}. While the proposed approaches also rely on monotonicity, they target the ATE which is not a cross-world quantity. Thus, any inference resulting from our strategies can, in principle, be falsified by a future experiment. In Section~\ref{sec:alternative_model}, we also show how our approach can be used in other IV models, without invoking cross-world assumptions.

\subsection{ATE identification as a missing data problem}
\label{sec:ate_identification}
For brevity, we write $\mu_t^{(a)} = \mathbb{E}[Y_i(a) \mid T_i = t]$ for the conditional potential outcome mean of type $t$ under treatment $a$. In the setting described in Section~\ref{sec:setup}, the ATE can be decomposed as
\begin{equation}
    \ATE = \pi_C \bigl(\mu_C^{(1)} - \mu_C^{(0)}\bigr) + \pi_A \bigl(\mu_A^{(1)} - \mu_A^{(0)}\bigr) + \pi_N \bigl(\mu_N^{(1)} - \mu_N^{(0)}\bigr).
\label{eq:ate_decomp}
\end{equation}

The fraction of each causal type $\pi_C, \pi_A, \pi_N$ and four of the six potential outcome means are identified: both complier means $\mu_C^{(0)}$ and $\mu_C^{(1)}$, the always-taker mean under treatment $\mu_A^{(1)}$, and the never-taker mean under no treatment $\mu_N^{(0)}$ (see \ref{app:identification}). The remaining two potential outcome means, denoted $\sensAT = \mu_A^{(0)}$ and $\sensNT = \mu_N^{(1)}$, are not identified.

This decomposition isolates the quantities that prevent point identification of the ATE under assumptions \eqref{ass:1}--\eqref{ass:5}. Building on this, we next present two simple approaches for conducting credible and informative inference on the ATE.

\subsection{Sensitivity analysis}
\label{sec:sensitivity}
We first approach the problem from a sensitivity analysis perspective. The two quantities that prevent point identification of the ATE are $\sensAT = \mu_A^{(0)}$ and $\sensNT = \mu_N^{(1)}$, which we treat as sensitivity parameters. The decision problem introduced in Section~\ref{sec:setup} requires us to determine whether inequality~\eqref{eq:criterion} holds. Rearranging~\eqref{eq:criterion} separately for each sensitivity parameter yields
\begin{equation}
    \sensAT < \frac{\pi_C \bigl(\mu_C^{(1)} - \mu_C^{(0)}\bigr) + \pi_A \mu_A^{(1)} + \pi_N \bigl(\sensNT - \mu_N^{(0)}\bigr) - x}{\pi_A} \equiv \thresholdAT,
    \label{eq:solve_a}
\end{equation}
\begin{equation}
    \sensNT > \frac{x - \pi_C \bigl(\mu_C^{(1)} - \mu_C^{(0)}\bigr) - \pi_A \bigl(\mu_A^{(1)} - \sensAT\bigr) + \pi_N \mu_N^{(0)}}{\pi_N} \equiv \thresholdNT.
    \label{eq:solve_b}
\end{equation}
Both inequalities are rearrangements of the single linear constraint $\ATE > x$, which defines a half-plane in the two-dimensional parameter space of $(\sensAT, \sensNT)$. By setting $\sensNT = \thresholdNT$, we obtain the boundary of this half plane denoted $\mathcal{B}$. Any parameter tuple $(\sensAT, \sensNT)$ lying above this line is sufficient to guarantee $\ATE > x$.

\begin{figure}[h]
\centering
\begin{tikzpicture}[>=stealth, shorten >=3pt, shorten <=3pt]
\def\cellw{3.0}
\def\cellh{2.2}
\def\xC{\cellw}
\def\xA{2*\cellw}
\def\xN{3*\cellw}
\def\yOne{\cellh}    %
\def\yZero{0}        %
\node[font=\bfseries] at (\xC, 1.5*\cellh+0.4) {Compliers};
\node[font=\bfseries] at (\xA, 1.5*\cellh+0.4) {Always-takers};
\node[font=\bfseries] at (\xN, 1.5*\cellh+0.4) {Never-takers};
\node[font=\bfseries, anchor=east] at (0.4*\cellw, \yOne)  {$\mathbb{E}[Y_i(1)]$};
\node[font=\bfseries, anchor=east] at (0.4*\cellw, \yZero) {$\mathbb{E}[Y_i(0)]$};
\node (a)      at (\xC, \yOne)  {$\mu_C^{(1)}$};
\node (c)      at (\xA, \yOne)  {$\mu_A^{(1)}$};
\node[circle, draw, inner sep=2pt] (thetaN) at (\xN, \yOne) {$\sensNT$};
\node (b)      at (\xC, \yZero) {$\mu_C^{(0)}$};
\node[circle, draw, inner sep=2pt] (thetaA) at (\xA, \yZero) {$\sensAT$};
\node (d)      at (\xN, \yZero) {$\mu_N^{(0)}$};
\draw[<->] (a.south)      -- (b.north);
\draw[dashed, <->] (c.south)      -- (thetaA.north);
\draw[dashed, <->] (thetaN.south) -- (d.north);
\draw[<->, bend left=30] (a.north east)  to (c.north west);
\draw[dashed, <->, bend left=35] (a.north east)  to (thetaN.north west);
\draw[dashed, <->, bend left=30] (c.north east)  to (thetaN.north west);
\draw[dashed, <->, bend right=30] (b.south east)      to (thetaA.south west);
\draw[<->, bend right=35] (b.south east)      to (d.south west);
\draw[dashed, <->, bend right=30] (thetaA.south east) to (d.south west);
\end{tikzpicture}
\caption{Identified quantities ($\mu_C^{(1)}$, $\mu_C^{(0)}$, $\mu_A^{(1)}$, $\mu_N^{(0)}$) and unidentified sensitivity parameters ($\sensAT$, $\sensNT$; circled) by compliance type and potential outcome. Vertical arrows indicate within-type comparisons and horizontal arrows indicate cross-type comparisons. Solid arrows indicate observed comparisons and dashed arrows indicate implied comparisons.}
\label{fig:roster}
\end{figure}
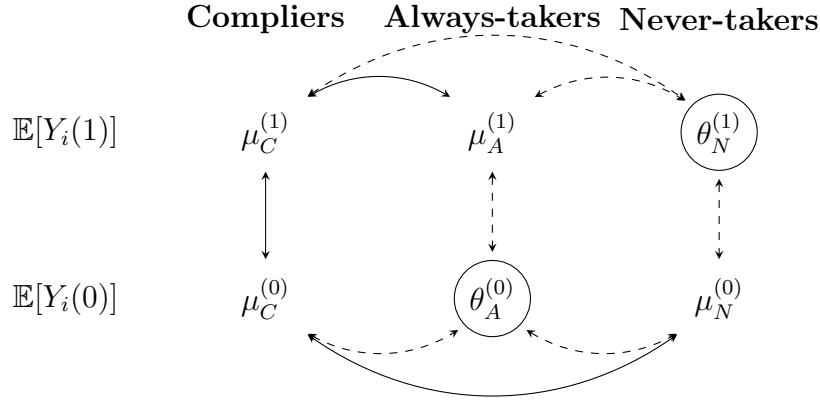

The decision problem thus reduces to the question of whether the sensitivity parameters lie above the boundary $\mathcal{B}$. Decision makers can assess this by judging the plausibility of the sensitivity parameter values on the boundary, i.e.\ $(\sensAT, \sensNT) \in \mathcal{B}$, either directly or in relation to identified quantities. We refer to quantities evaluated on $\mathcal{B}$ as \emph{implied quantities}, and to contrasts between them and identified quantities as \emph{implied comparisons}. Figure~\ref{fig:roster} illustrates possible such comparisons.

Within-type comparisons (vertical arrows in Figure~\ref{fig:roster}) relate the implied value of a sensitivity parameter to an identified quantity for the same compliance type. For example, one can compute the implied additive effect ($\thresholdNT - \mu_N^{(0)}$) or risk ratio ($\thresholdNT / \mu_N^{(0)}$) among never-takers. Since $\thresholdNT$ depends on $\sensAT$, these implied quantities vary along $\mathcal{B}$. Cross-type comparisons (horizontal arrows in Figure~\ref{fig:roster}) relate implied quantities to identified quantities for other compliance types, giving a notion of outcome heterogeneity. For example, $\thresholdNT - \mu_C^{(1)}$ captures implied heterogeneity between never-takers and compliers.

These comparisons can also be expressed relative to identified counterparts: the implied additive effect among never-takers ($\thresholdNT - \mu_N^{(0)}$) can be contrasted with the identified effect among compliers ($\mu_C^{(1)} - \mu_C^{(0)}$), or the implied outcome heterogeneity between never-takers and compliers ($\thresholdNT - \mu_C^{(1)}$) can be benchmarked against the identified heterogeneity between always-takers and compliers ($\mu_A^{(1)} - \mu_C^{(1)}$). 

Additional context for these comparisons can be obtained from baseline covariates. For a baseline covariate $L_i$, the mean among each compliance type is identified under assumptions \eqref{ass:1}--\eqref{ass:5} \citep{abadie_semiparametric_2003, marbach_profiling_2020}, provided that independence \eqref{ass:2} also holds for $L_i$, which is the case when the instrument is randomized. If compliance types are similar with respect to baseline covariates, large implied differences in treatment effects or outcomes between them are arguably less plausible than when these covariates differ substantially. For additional considerations that may inform comparisons between the potential outcome means of different strata, we refer to \cite{kowalski_how_2023}.

The analysis simplifies considerably under one-sided non-compliance. This commonly arises, for example, in randomized controlled trials (RCTs) where participants in the control arm are unable to access a novel treatment. Without loss of generality, suppose $P(A_i = 1 \mid Z_i = 0) = 0$, which is empirically testable and implies both $\pi_A = 0$ and monotonicity~\eqref{ass:5}. This means that always-takers are absent, $\sensAT$ is no longer relevant, and the LATE becomes equal to the average treatment effect on the treated (ATT), defined as $\Psi_{\text{ATT}} = E[Y_i(1) - Y_i(0) \mid A_i = 1]$. The two-dimensional problem then collapses to a single inequality: the decision criterion $\ATE > x$ holds if and only if
\begin{equation}
    \sensNT > \frac{x - \pi_C \bigl(\mu_C^{(1)} - \mu_C^{(0)}\bigr) + \pi_N \mu_N^{(0)}}{\pi_N} \equiv \thresholdNTstar.
    \label{eq:solve_simp}
\end{equation}
In this case, a single scalar threshold $\thresholdNTstar$ fully characterizes the decision problem: $\ATE > x$ if and only if $\sensNT > \thresholdNTstar$. This threshold can again be interpreted in relation to identified quantities, as illustrated in Figure~\ref{fig:roster}, except that the always-taker column drops out.

In Section~\ref{sec:examples}, we apply these results to examples with one-sided and two-sided non-compliance, and derive sufficient assumptions to establish the sign of the ATE.

\subsection{Improved bounds}
\label{sec:bounds}
The sensitivity analysis in Section~\ref{sec:sensitivity} addresses the question of how extreme the unidentified quantities are allowed to be for the decision problem to be resolved. As previously proposed by \cite{chen_going_2018}, researchers can also use their background knowledge to directly restrict the plausible range of the sensitivity parameters and propagate these restrictions into bounds on the ATE. For binary outcomes, imputing the most extreme feasible values for $\sensAT$ and $\sensNT$ yields bounds that coincide with the sharp bounds derived by \cite{balke_bounds_1997} (see \ref{app:bp}). This gives intuition for why assumption-free bounds are often wide: they are driven by extreme but typically implausible values of the missing potential outcomes. Restricting $\sensAT$ and $\sensNT$ to a subset of their feasible range narrows the bounds.

Formally, suppose the researcher is willing to assume that $\sensAT \in [\ell_A, u_A]$ and $\sensNT \in [\ell_N, u_N]$, where these intervals encode the restrictions they are confident to impose. Let $\ATE(\sensAT, \sensNT)$ denote the value of the average treatment effect as a function of the sensitivity parameters, as given by equation~\eqref{eq:ate_decomp}. Since $\ATE(\sensAT, \sensNT)$ is decreasing in $\sensAT$ and increasing in $\sensNT$, the improved bounds are
\begin{equation}
    \ATE \in \left[\ATE(u_A, \ell_N),\ \ATE(\ell_A, u_N)\right],
    \label{eq:bounds}
\end{equation}
and are obtained by evaluating $\ATE(\sensAT, \sensNT)$ at the relevant corners of the feasible set $[\ell_A, u_A] \times [\ell_N, u_N]$.

The intervals $[\ell_A, u_A]$ and $[\ell_N, u_N]$ should be constructed in a principled way, based on substantive background knowledge. One approach is to bound the sensitivity parameters directly. Another is to restrict quantities that depend on the sensitivity parameters, analogous to the threshold comparisons in Section~\ref{sec:sensitivity}. For example, we can bound the treatment effect among never-takers, $\sensNT - \mu_N^{(0)}$, or the outcome heterogeneity between never-takers and compliers, $\sensNT - \mu_C^{(1)}$. Any bound on such a derived quantity translates directly into a bound on the sensitivity parameter by solving the expression for $\sensNT$, and the same logic applies to $\sensAT$ and to the other comparisons described in Section~\ref{sec:sensitivity}. When the researcher imposes bounds on more than one such quantity, each restriction defines a feasible interval for the relevant sensitivity parameter, and the overall feasible set is the intersection of these intervals. This provides flexibility to researchers: assumptions can be stated on whichever scale is most natural and combined freely, with the implied bounds on the ATE always obtained by evaluating at the corners of the resulting feasible set.

\subsection{Inference}
In practice, the sensitivity thresholds and bounds must be estimated from finite-sample data. Both the boundary line $\mathcal{B}$ in the two-sided non-compliance case and the scalar threshold $\thresholdNTstar$ in the one-sided case are linear combinations of the identified quantities, each of which is a smooth function of the conditional means of the observed data (see \ref{app:identification}), provided the relevant compliance-type probabilities are bounded away from zero. Plugging the sample analogues of these conditional means into expressions \eqref{eq:solve_a}--\eqref{eq:solve_simp} yields plug-in estimators $\thresholdNTstarhat$ of the threshold in the one-sided case, and $\hat{\mathcal{B}}$ of every point on the boundary in the two-sided case. 

The improved bounds in Section~\ref{sec:bounds} are obtained by evaluating $\Psi_{\text{ATE}}(\sensAT, \sensNT)$ at the corners of the feasible set $[\ell_A, u_A] \times [\ell_N, u_N]$. Since $\Psi_{\text{ATE}}(\sensAT, \sensNT)$ is again a linear combination of the identified quantities for any fixed $(\sensAT, \sensNT)$, the same argument applies and the corresponding plug-in estimators of the upper and lower bound are consistent.

The nonparametric bootstrap consistently estimates the sampling distributions of the thresholds, boundary points, and fixed-corner bounds under standard iid sampling and positivity conditions, since these are smooth functions of empirical conditional means and probabilities \citep{efron_bootstrap_1986}. In each bootstrap sample, the identified quantities are estimated and the threshold, boundary, or bounds are recomputed. Confidence intervals (or pointwise confidence bands in the case of $\hat{\mathcal{B}}$) are then constructed from the resulting empirical distribution across bootstrap samples.

A disadvantage of the plug-in approach is that, in finite samples, the estimated observed-data distribution can violate the testable restrictions of the IV model, yielding estimates that are incompatible with the model, such as $\hat{\pi}_C < 0$ or, for a binary outcome, $\hat{\mu}_C^{(0)} \notin [0,1]$. We recommend checking and reporting whether this occurs in the original sample and in what proportion of bootstrap samples. A violation in the original sample may indicate that assumptions \eqref{ass:1}--\eqref{ass:5} fail, which can be assessed with a formal test of the model's testable restrictions \citep{kitagawa_test_2015, wang_falsification_2017}. However, violations can also arise when the true distribution lies in the model, particularly with weak instruments, rare outcomes, or small samples. If such violations are frequent across bootstrap samples, the bootstrap intervals may be unreliable. For discrete outcomes, the finite-sample confidence regions of \citet{guo_chernoff-type_2021} provide an alternative that respects the model restrictions by construction.

Alternatively, Bayesian inference can be conducted using the transparent parametrization suggested by \cite{gustafson_model_2005} and \cite{richardson_transparent_2011}. For binary outcomes, a posterior on the identified quantities can be obtained using the procedures described in Sections~3.1 and~4.3 of \cite{richardson_transparent_2011}, which rely solely on (straight) Monte-Carlo procedures. Since the threshold $\thresholdNTstar$ and the boundary $\mathcal{B}$ are functions of the identified quantities alone, their posterior distributions follow directly. For a Bayesian sensitivity analysis, the posterior of the ATE can be evaluated at fixed values of the sensitivity parameters, yielding a posterior of the ATE as a function of $(\sensAT, \sensNT)$ \citep[Section~3.2]{richardson_transparent_2011}. Researchers may also specify a prior on $(\sensAT, \sensNT)$ based on substantive knowledge, possibly informed by comparisons in Figure~\ref{fig:roster}, to obtain a marginal posterior of the ATE.

The proposed frequentist plug-in and Bayesian procedures are implemented in the accompanying R package \ivate.

\subsection{Single-world instrumental variable model}
\label{sec:alternative_model}
So far, we have focused on an IV model that requires monotonicity. We now consider a different IV model that replaces monotonicity with a different homogeneity assumption and allows for point identification of the ATT, rather than the LATE.

Following \citet{hernan_causal_2019} (Technical Point~16.3), point identification of the ATT may also be obtained from the IV assumptions \eqref{ass:1}--\eqref{ass:4} and \eqref{ass:3} together with a homogeneity assumption stating that the instrument does not modify the ATT,
\begin{equation}
    E[Y_i(1) - Y_i(0) \mid A_i = 1, Z_i = 1]
    =
    E[Y_i(1) - Y_i(0) \mid A_i = 1, Z_i = 0].
    \label{eq:homogeneity}
\end{equation}
Unlike in the model considered so far, monotonicity is not required.

Under assumptions \eqref{ass:1}--\eqref{ass:4}, \eqref{ass:3} and \eqref{eq:homogeneity}, the ATT is point identified. While the assumptions are insufficient for point identification of the ATE, it can again be decomposed into identified and unidentified components:
\begin{equation}
    \Psi_{\mathrm{ATE}}
    =
    P(A_i = 1)\,\Psi_{\mathrm{ATT}}
    +
    P(A_i = 0)\,E[Y_i(1)-Y_i(0)\mid A_i = 0],
    \label{eq:ate_decomp_att}
\end{equation}
where all quantities except $\theta^{(1)} = E[Y_i(1)\mid A_i = 0]$ are identified.

The strategies introduced in Sections \ref{sec:sensitivity} and \ref{sec:bounds} apply directly in this setting. Treating $\theta^{(1)}$ as a sensitivity parameter yields a one-dimensional sensitivity analysis analogous to the one-sided non-compliance case in Section \ref{sec:sensitivity}. In particular, the decision criterion $\Psi_{\mathrm{ATE}} > x$ reduces to a single threshold condition on $\theta^{(1)}$. Likewise, imposing restrictions on $\theta^{(1)}$ or on derived contrasts involving $\theta^{(1)}$ yields corresponding bounds on the ATE.

This illustrates that the proposed framework generalizes naturally to any model that admits a decomposition of the ATE into identified and unidentified components. It is not limited to the IV model discussed in this section, but extends to other methods that target the ATT (e.g. difference-in-differences) or related parameters.

\section{Examples}
\label{sec:examples}
To illustrate the value of the proposed approach, we present two real-data examples. The cell counts required to reproduce all results are given in \ref{app:counts}. Bootstrapped 95\% confidence intervals are given in parentheses. In these examples, neither the original nor any of the bootstrapped samples violates any of the model's testable restrictions.

\subsection{Example 1: Supplemental MRI screening for breast cancer}

As a first example, consider \cite{bakker_supplemental_2019}, who conducted an RCT to study the effect of supplemental MRI screening on breast cancer incidence among women with extremely dense breast tissue ($n = 40,373$). Women in the treatment group were invited to undergo MRI screening in addition to the standard mammography received by all participants. Non-compliance was one-sided: 41\% of the women in the treatment group remained unexposed to the intervention, while none of the women in the control group were exposed (i.e. no always-takers exist).

Estimation of the LATE yields $\hat{\Psi}_{\mathrm{LATE}} = -0.0042\ (-0.0064, -0.0019)$. Without restrictions on the sensitivity parameters, the estimated plug-in bounds on the ATE, which coincide with the Balke--Pearl bounds under our model, are $\hat{\Psi}_{\mathrm{ATE}} \in [-0.0045\ (-0.0053, -0.0035),$ $0.402\ (0.391, 0.412)]$. Suppose a decision maker wants to determine whether the ATE is negative---that is, whether the treatment reduces cancer incidence. Among compliers, $\hat{\Psi}_{\mathrm{LATE}}$ suggests the treatment is beneficial, but the bounds show that without additional assumptions the data are consistent with a wide range of effects at the population level, with the majority of this range implying harm. The question of whether the treatment is beneficial at the population level therefore cannot be resolved without additional information.

Applying the decomposition~\eqref{eq:ate_decomp} to the data yields
\[
\hat{\Psi}_{\mathrm{ATE}} = \underbrace{0.593}_{\hat{\pi}_C} \bigl(\underbrace{0.00084}_{\hat{\mu}_C^{(1)}} - \underbrace{0.0051}_{\hat{\mu}_C^{(0)}}\bigr) + \underbrace{0.407}_{\hat{\pi}_N} \bigl(\sensNT - \underbrace{0.0049}_{\hat{\mu}_N^{(0)}}\bigr).
\]

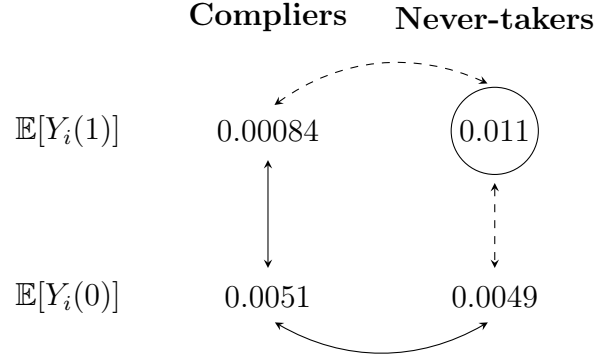
\begin{figure}[t]
\centering
\begin{tikzpicture}[>=stealth, shorten >=3pt, shorten <=3pt]
\def\cellw{3.0}
\def\cellh{2.2}
\def\xC{\cellw}
\def\xN{2*\cellw}
\def\yOne{\cellh}    %
\def\yZero{0}        %
\node[font=\bfseries] at (\xC, 1.5*\cellh+0.4) {Compliers};
\node[font=\bfseries] at (\xN, 1.5*\cellh+0.4) {Never-takers};
\node[font=\bfseries, anchor=east] at (0.4*\cellw, \yOne)  {$\mathbb{E}[Y_i(1)]$};
\node[font=\bfseries, anchor=east] at (0.4*\cellw, \yZero) {$\mathbb{E}[Y_i(0)]$};
\node (a)      at (\xC, \yOne)  {$0.00084$};
\node[circle, draw, inner sep=2pt] (thetaN) at (\xN, \yOne) {$0.011$};
\node (b)      at (\xC, \yZero) {$0.0051$};
\node (d)      at (\xN, \yZero) {$0.0049$};
\draw[<->] (a.south)      -- (b.north);
\draw[dashed, <->] (thetaN.south) -- (d.north);
\draw[dashed, <->, bend left=30] (a.north)  to (thetaN.north);
\draw[<->, bend right=30] (b.south)      to (d.south);
\end{tikzpicture}
\caption{Estimated identified potential outcome means and the estimated threshold $\thresholdNTstarhat$ for the sensitivity parameter $\sensNT$ for \cite{bakker_supplemental_2019}. The decision criterion $\ATE < 0$ holds if and only if $\sensNT$ lies below the threshold. Arrows are defined as in Figure~\ref{fig:roster}.}
\label{fig:example_bakker}
\end{figure}

Because non-compliance is one-sided, point identification of the ATE is precluded only by ignorance of $\sensNT = \mu_N^{(1)}$. The upper bound of $0.402$ reported above corresponds to the extreme scenario in which $\sensNT = 1$, implying that every never-taker would develop cancer if exposed to the MRI screening. The implausibility of this scenario highlights the potential for additional assumptions to improve the inference on the ATE. The decision criterion $\ATE < 0$ holds if and only if $\sensNT < \thresholdNTstar$. Figure~\ref{fig:example_bakker} places this threshold in relation to the identified quantities, and Table~\ref{tab:iv_sensitivity} summarizes implied comparisons and bounds under different assumptions.

The threshold obtained (see (1) in Table~\ref{tab:iv_sensitivity}) implies that the treatment is beneficial for all values of $\sensNT$ smaller than $\thresholdNTstarhat = 0.011\ (0.009, 0.013)$. At this threshold, supplemental MRI would increase cancer incidence by $0.006\ (0.003, 0.009)$ among never-takers, corresponding to a $2.26$-fold increase on the risk ratio scale. Moreover, the implied outcome heterogeneity between never-takers and compliers under treatment, $\thresholdNTstarhat - \hat{\mu}_C^{(1)} = 0.010\ (0.007, 0.013)$, is large relative to the observed outcome heterogeneity under no treatment, which is very close to zero, $\hat{\mu}_C^{(0)} - \hat{\mu}_N^{(0)} = 0.0002\ (-0.0043, 0.0041)$. If one is willing to reject the plausibility of any of these implied values, one can confidently conclude that the decision criterion is met.

Alternatively, a decision maker might impose restrictions on the sensitivity parameters to obtain tighter bounds, for example, by restricting the treatment effect among never-takers. If one is willing to assume that the intervention increases cancer incidence by at most a factor of $1.2$ among never-takers, the resulting bounds are $\hat{\Psi}_{\mathrm{ATE}} \in [-0.0045\ (-0.0053, -0.0035),$ $-0.0021\ (-0.0036, -0.0006)]$ (see (2) in Table~\ref{tab:iv_sensitivity}). Under this assumption, the bounds exclude zero, thereby resolving the decision problem.

\subsection{Example 2: Public health insurance and health care utilization}

As a second example, consider \cite{finkelstein_oregon_2012}, who conducted an RCT to study the effect of access to public health insurance on health care utilization ($n =$ 23,492). Participants assigned to the treatment group were offered access to Medicaid. Non-compliance was two-sided: 57\% of participants in the treatment group did not enroll, while 13\% of participants in the control group obtained coverage through other means. Suppose a decision maker wants to determine whether the intervention was on average beneficial, i.e.\ whether $\ATE > 0$.

Estimation of the LATE yields $\hat{\Psi}_{\mathrm{LATE}} = 0.193\ (0.153, 0.237)$, and the unrestricted plug-in bounds on the ATE are $\hat{\Psi}_{\mathrm{ATE}} \in [-0.288\ (-0.299, -0.276),\; 0.416\ (0.405, 0.426)]$. As in the first example, the data alone are insufficient to resolve the decision problem without further assumptions.

Applying the decomposition~\eqref{eq:ate_decomp} to the data yields
\[
\hat{\Psi}_{\mathrm{ATE}} = \underbrace{0.30}_{\hat{\pi}_C} \bigl(\underbrace{0.70}_{\hat{\mu}_C^{(1)}} - \underbrace{0.51}_{\hat{\mu}_C^{(0)}}\bigr) + \underbrace{0.13}_{\hat{\pi}_A} \bigl(\underbrace{0.81}_{\hat{\mu}_A^{(1)}} - \sensAT\bigr) + \underbrace{0.57}_{\hat{\pi}_N} \bigl(\sensNT - \underbrace{0.56}_{\hat{\mu}_N^{(0)}}\bigr).
\]

\begin{figure}[t]
\centering
\begin{tikzpicture}[>=stealth, shorten >=3pt, shorten <=3pt]
\def\cellw{4.2}
\def\cellh{2.2}
\def\xC{\cellw}
\def\xA{2*\cellw}
\def\xN{3*\cellw}
\def\yOne{\cellh}    %
\def\yZero{0}        %
\node[font=\bfseries] at (\xC, 1.5*\cellh+0.8) {Compliers};
\node[font=\bfseries] at (\xA, 1.5*\cellh+0.8) {Always-takers};
\node[font=\bfseries] at (\xN, 1.5*\cellh+0.8) {Never-takers};
\node[font=\bfseries, anchor=east] at (0.55*\cellw, \yOne)  {$\mathbb{E}[Y_i(1)]$};
\node[font=\bfseries, anchor=east] at (0.55*\cellw, \yZero) {$\mathbb{E}[Y_i(0)]$};
\node (a)      at (\xC, \yOne)  {$0.70$};
\node (c)      at (\xA, \yOne)  {$0.81$};
\node[rectangle, rounded corners=3pt, draw, inner sep=3pt] (thetaN) at (\xN, \yOne) {$\sensNT > 0.28 + 0.22\,\sensAT$};
\node (b)      at (\xC, \yZero) {$0.51$};
\node[rectangle, rounded corners=3pt, draw, inner sep=3pt] (thetaA) at (\xA, \yZero) {$\sensAT < -1.24 + 4.46\,\sensNT$};
\node (d)      at (\xN, \yZero) {$0.56$};
\draw[<->] (a.south)      -- (b.north);
\draw[dashed, <->] (c.south)      -- (thetaA.north);
\draw[dashed, <->] (thetaN.south) -- (d.north);
\draw[<->, bend left=25] (a.north)  to (c.north);
\draw[dashed, <->, bend left=30] (a.north)  to (thetaN.north);
\draw[dashed, <->, bend left=25] (c.north)  to (thetaN.north);
\draw[dashed, <->, bend right=25] (b.south)      to (thetaA.south);
\draw[<->, bend right=30] (b.south)      to (d.south);
\draw[dashed, <->, bend right=25] (thetaA.south) to (d.south);
\end{tikzpicture}
\caption{Estimated identified potential outcome means and the estimated boundary $\hat{\mathcal{B}}$ for the sensitivity parameters $\sensAT$ and $\sensNT$ for \cite{finkelstein_oregon_2012}. The entry for each sensitivity parameter gives the condition on it, given the other, under which the decision criterion $\ATE > 0$ holds; the two conditions are equivalent. Arrows are defined as in Figure~\ref{fig:roster}.}
\label{fig:example_finkelstein}
\end{figure}
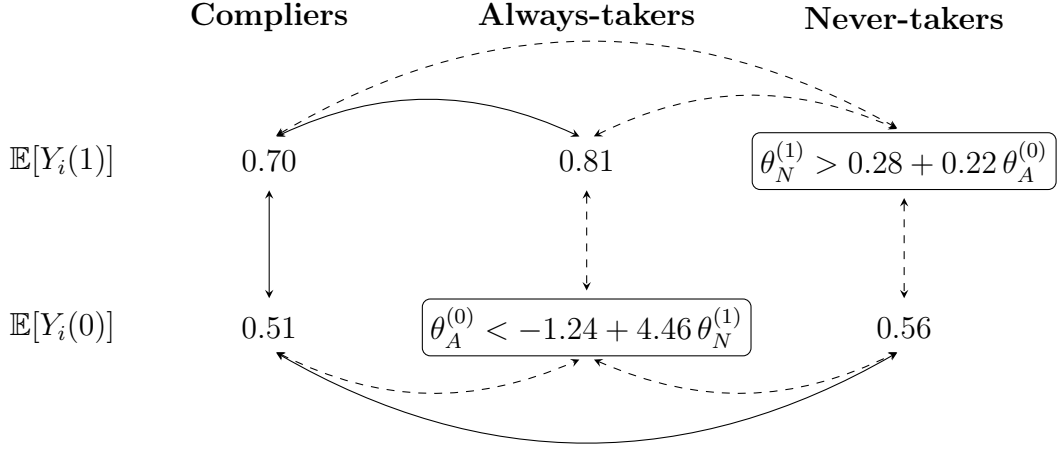

Because non-compliance is two-sided, point identification of the ATE is precluded by our ignorance of both $\sensNT = \mu_N^{(1)}$ and $\sensAT = \mu_A^{(0)}$. Figure~\ref{fig:example_finkelstein} illustrates the derived boundary $\mathcal{B}$ in relation to the identified quantities, and Table~\ref{tab:iv_sensitivity} summarizes implied comparisons and bounds under different assumptions.

\begin{table}[ht]
\centering
\scriptsize
\renewcommand{\arraystretch}{1.5}
\begin{tabular}{@{}ccp{2.2cm}cp{3.4cm}p{5.5cm}@{}}
\toprule
 & Ex. & Restrictions & Bounds & Threshold & Implied effect \\
\midrule
(1) & I & -- & $[-0.004,\; 0.40]$ & $\theta_N^{(1)} < 0.011$ & $\theta_N^{(1)} - \hat{\mu}_N^{(0)} < 0.006$ \\
(2) & I & $\theta_N^{(1)} / \hat{\mu}_N^{(0)} \leq 1.2$ & $[-0.004,\; -0.002]$ & -- & -- \\
\midrule
(3) & II & -- & $[-0.29,\; 0.42]$ & $\theta_A^{(0)} < -1.24 + 4.46 \cdot \theta_N^{(1)}$, $\theta_N^{(1)} > 0.28 + 0.22 \cdot \theta_A^{(0)}$ & $\hat{\mu}_A^{(1)} - \theta_A^{(0)} > -0.44 - 4.46 \cdot (\theta_N^{(1)} - \hat{\mu}_N^{(0)})$, $\theta_N^{(1)} - \hat{\mu}_N^{(0)} > -0.099 - 0.22 \cdot (\hat{\mu}_A^{(1)} - \theta_A^{(0)})$ \\
(4) & II & $\theta_A^{(0)} = 1$ & $[-0.29,\; 0.29]$ & $\theta_N^{(1)} > 0.50$ & $\theta_N^{(1)} - \hat{\mu}_N^{(0)} > -0.056$ \\
(5) & II & $\hat{\mu}_A^{(1)} - \theta_A^{(0)} \geq 0$ & $[-0.26,\; 0.42]$ & $\theta_N^{(1)} > 0.46$ & $\theta_N^{(1)} - \hat{\mu}_N^{(0)} > -0.099$ \\
(6) & II & $|\theta_N^{(1)} - \hat{\mu}_C^{(1)}| \leq |\hat{\mu}_N^{(0)} - \hat{\mu}_C^{(0)}|$, $|\theta_A^{(0)} - \hat{\mu}_C^{(0)}| \leq |\hat{\mu}_A^{(1)} - \hat{\mu}_C^{(1)}|$ & $[0.14,\; 0.22]$ & -- & -- \\
\bottomrule
\end{tabular}
\caption{Sensitivity thresholds and bounds for different examples and assumptions.
          Example~I is \cite{bakker_supplemental_2019} with decision criterion $\ATE < 0$;
          Example~II is \cite{finkelstein_oregon_2012} with decision criterion $\ATE > 0$.}
\label{tab:iv_sensitivity}
\end{table}

The two-sided case is more complex than the one-sided case because the two sensitivity parameters can vary simultaneously. Rather than a single threshold, the decision criterion $\ATE > 0$ is characterized by a boundary function relating $\sensAT$ and $\sensNT$ (see (3) in Table~\ref{tab:iv_sensitivity}). To assess whether this boundary can be rejected, we propose a simple two-step strategy.

In the first step, one fixes either sensitivity parameter at its worst-case value (here, \(\sensAT = 1\) or \(\sensNT = 0\)) and evaluates the implied threshold for the remaining parameter. For continuous outcomes, a worst-case bound must first be defined. Setting $\sensAT = 1$ yields an implied threshold of $\hat{\tau}_N^{(1)} = 0.501\ (0.483, 0.518)$, corresponding to an implied treatment effect of $\hat{\tau}_N^{(1)} - \hat{\mu}_N^{(0)} = -0.056\ (-0.079, -0.036)$ among never-takers (see (4) in Table~\ref{tab:iv_sensitivity}). Moreover, the implied outcome heterogeneity between never-takers and compliers under treatment, $\hat{\tau}_N^{(1)} - \hat{\mu}_C^{(1)} = -0.202\ (-0.237, -0.171)$, is more than four times as pronounced as the observed outcome heterogeneity under no treatment, $\hat{\mu}_C^{(0)} - \hat{\mu}_N^{(0)} = -0.047\ (-0.095, -0.005)$. If one is willing to reject the plausibility of these values, the decision problem is resolved without any further assumptions.

If such rejection is not possible, one may consider in a second step whether one of the two parameters can be credibly restricted to a subset of its feasible range. Taking the worst-case value within this restricted set then yields a tighter threshold for the remaining parameter. For instance, if one is willing to assume that the treatment effect among always-takers is non-negative, i.e.\ $\mu_A^{(1)} - \sensAT \geq 0$, the implied threshold for $\sensNT$ tightens to $\hat{\tau}_N^{(1)} = 0.458\ (0.440, 0.476)$, with a corresponding implied treatment effect of $\hat{\tau}_N^{(1)} - \hat{\mu}_N^{(0)} = -0.099\ (-0.122, -0.079)$ among never-takers (see (5) in Table~\ref{tab:iv_sensitivity}). If one is comfortable rejecting these values, the decision problem is again resolved.

Alternatively, one can obtain tighter bounds by restricting the plausible range of the sensitivity parameters. For example, one may restrict outcome heterogeneity across compliance types under the unobserved treatment condition: assuming that never-takers and compliers differ under treatment by at most as much as under no treatment, $|\sensNT - \mu_C^{(1)}| \leq |\mu_N^{(0)} - \mu_C^{(0)}|$, and analogously for always-takers, $|\sensAT - \mu_C^{(0)}| \leq |\mu_A^{(1)} - \mu_C^{(1)}|$, yields bounds of $[0.139\ (0.111, 0.163),\; 0.220\ (0.185, 0.261)]$, which exclude zero and resolve the decision problem (see (6) in Table~\ref{tab:iv_sensitivity}).

\section{Discussion}
\label{sec:discussion}
We studied two complementary strategies for inference on ATE in IV settings: (i) a sensitivity analysis that derives thresholds on the latent quantities preventing point identification, and (ii) a related approach that translates direct restrictions on these quantities into improved bounds on the ATE. We illustrated both strategies using two real-data examples, demonstrating how they can support decision-making in practice.

When the ATE is of primary interest, standard IV methods often fail to provide a conclusive answer: a point-identified LATE gives only a partial answer to the substantive research question, a point-identified ATE is rarely attainable under plausible assumptions, and assumption-free bounds on the ATE are often too wide to be informative. Our framework addresses this tension, enabling both researchers and consumers of evidence to assess whether the available data and credible assumptions are sufficient to resolve a decision problem that depends on the ATE. Making explicit the assumptions necessary for a given conclusion to hold adds relevant information and may reduce the risk of misinterpretation, which is important when results are used to support practical decisions \citep{locher_interpretational_2025}. We therefore argue that reporting sensitivity thresholds would be a valuable complement to existing guidelines for reporting of IV analyses \citep{swanson_commentary_2013}.

The proposed strategies are straightforward to implement, which will facilitate their adoption in applied research. The sensitivity parameters have a natural interpretation as unobserved potential outcome means, and the derived thresholds can be expressed relative to identified quantities. This makes the results easy to communicate. Conceptually, our approach is closely related to sensitivity analyses for violations of identification assumptions, such as the E-value for unobserved confounding \citep{vanderweele_sensitivity_2017}, which represents a threshold on the strength of unmeasured confounding required to explain away an effect. Our results also involve thresholds, but unlike those in \citet{vanderweele_sensitivity_2017}, they concern the magnitude of causal estimands, such as the potential outcome mean among never-takers under treatment, rather than the extent of unmeasured confounding. This distinction is important because our thresholds are not subject to the critique that they fail to account for the prevalence of the unmeasured common causes \citep{ioannidis_limitations_2019, sjolander_are_2022}.

The proposed methods have several limitations. First, while they accommodate both binary and continuous outcomes, they are restricted to settings with binary instruments and treatments. Second, we focus on an IV model that relies on the monotonicity assumption. While in one-sided non-compliance settings this assumption holds by design, it is not consistently testable in the presence of two-sided non-compliance and must therefore be justified on substantive grounds. However, we have shown that our strategy generalizes naturally to other IV models that permit a decomposition of the ATE into identified and unidentified components. Third, in settings with two-sided non-compliance, the presence of two sensitivity parameters requires joint assessment, which may be more demanding than in the one-sided case. Although we propose practical strategies to simplify this evaluation, the resulting analysis is inherently more involved.

The question of which estimand to target in IV analyses has been subject to many debates \citep{deaton_instruments_2010, heckman_comparing_2010, imbens_better_2010-1, imbens_causality_2022}. We believe our contribution allows researchers, at least in settings with binary instrument and treatment, to conduct informative inference on the estimand of interest without having to sacrifice the credibility of assumptions.

\section*{Acknowledgments}
Claude Opus models (Anthropic), up to and including Claude Opus 5.5, were used for data analysis, development of the accompanying software, creation of figures, checking of proofs, and improving the clarity of writing. The authors reviewed, verified, and edited all AI-assisted output. The conceptual ideas are by the authors, who take full responsibility for the content of this paper.

We thank Peter Hull, participants of the European Causal Inference Meeting 2026, and seminar participants at Brown University, Harvard University, and the University of Massachusetts Amherst for helpful comments. Luca Locher is grateful to Leora Sarvet for inspiring discussions that motivated the project.

\newpage
\bibliographystyle{plainnat}
\bibliography{references}

\begin{thebibliography}{28}
\providecommand{\natexlab}[1]{#1}
\providecommand{\url}[1]{\texttt{#1}}
\expandafter\ifx\csname urlstyle\endcsname\relax
  \providecommand{\doi}[1]{doi: #1}\else
  \providecommand{\doi}{doi: \begingroup \urlstyle{rm}\Url}\fi

\bibitem[Abadie(2003)]{abadie_semiparametric_2003}
Alberto Abadie.
\newblock Semiparametric instrumental variable estimation of treatment response models.
\newblock \emph{Journal of econometrics}, 113\penalty0 (2):\penalty0 231--263, 2003.
\newblock Publisher: Elsevier.

\bibitem[Angrist et~al.(1996)Angrist, Imbens, and Rubin]{angrist_identification_1996}
Joshua~D. Angrist, Guido~W. Imbens, and Donald~B. Rubin.
\newblock Identification of {Causal} {Effects} {Using} {Instrumental} {Variables}.
\newblock \emph{Journal of the American Statistical Association}, 91\penalty0 (434):\penalty0 444--455, June 1996.
\newblock ISSN 0162-1459.
\newblock \doi{10.1080/01621459.1996.10476902}.
\newblock URL \url{https://www.tandfonline.com/doi/abs/10.1080/01621459.1996.10476902}.
\newblock Publisher: ASA Website \_eprint: https://www.tandfonline.com/doi/pdf/10.1080/01621459.1996.10476902.

\bibitem[Bakker et~al.(2019)Bakker, Lange, Pijnappel, Mann, Peeters, Monninkhof, Emaus, Loo, Bisschops, Lobbes, Jong, Duvivier, Veltman, Karssemeijer, Koning, Diest, Mali, Bosch, Veldhuis, and Gils]{bakker_supplemental_2019}
Marije~F. Bakker, Stéphanie V.~de Lange, Ruud~M. Pijnappel, Ritse~M. Mann, Petra H.~M. Peeters, Evelyn~M. Monninkhof, Marleen~J. Emaus, Claudette~E. Loo, Robertus H.~C. Bisschops, Marc B.~I. Lobbes, Matthijn D. F.~de Jong, Katya~M. Duvivier, Jeroen Veltman, Nico Karssemeijer, Harry J.~de Koning, Paul J.~van Diest, Willem P. T.~M. Mali, Maurice A. A. J. van~den Bosch, Wouter~B. Veldhuis, and Carla H.~van Gils.
\newblock Supplemental {MRI} {Screening} for {Women} with {Extremely} {Dense} {Breast} {Tissue}.
\newblock \emph{New England Journal of Medicine}, 381\penalty0 (22):\penalty0 2091--2102, November 2019.
\newblock ISSN 0028-4793.
\newblock \doi{10.1056/NEJMoa1903986}.
\newblock URL \url{https://www.nejm.org/doi/full/10.1056/NEJMoa1903986}.
\newblock Publisher: Massachusetts Medical Society \_eprint: https://www.nejm.org/doi/pdf/10.1056/NEJMoa1903986.

\bibitem[Balke and Pearl(1997)]{balke_bounds_1997}
Alexander Balke and Judea Pearl.
\newblock Bounds on {Treatment} {Effects} {From} {Studies} {With} {Imperfect} {Compliance}.
\newblock \emph{Journal of the American Statistical Association}, 92\penalty0 (439):\penalty0 1171--1176, 1997.
\newblock ISSN 0162-1459.
\newblock \doi{10.2307/2965583}.
\newblock URL \url{https://www.jstor.org/stable/2965583}.
\newblock Publisher: [American Statistical Association, Taylor \& Francis, Ltd.].

\bibitem[Chen et~al.(2018)Chen, Flores, and Flores-Lagunes]{chen_going_2018}
Xuan Chen, Carlos~A. Flores, and Alfonso Flores-Lagunes.
\newblock Going beyond {LATE}: {Bounding} {Average} {Treatment} {Effects} of {Job} {Corps} {Training}.
\newblock \emph{The Journal of Human Resources}, 53\penalty0 (4):\penalty0 1050--1099, 2018.
\newblock ISSN 0022-166X.
\newblock URL \url{https://www.jstor.org/stable/26533541}.
\newblock Publisher: [University of Wisconsin Press, Board of Regents of the University of Wisconsin System].

\bibitem[Deaton(2010)]{deaton_instruments_2010}
Angus Deaton.
\newblock Instruments, randomization, and learning about development.
\newblock \emph{Journal of economic literature}, 48\penalty0 (2):\penalty0 424--455, 2010.
\newblock Publisher: American Economic Association.

\bibitem[Efron and Tibshirani(1986)]{efron_bootstrap_1986}
Bradley Efron and Robert Tibshirani.
\newblock Bootstrap methods for standard errors, confidence intervals, and other measures of statistical accuracy.
\newblock \emph{Statistical science}, pages 54--75, 1986.
\newblock URL \url{https://www.jstor.org/stable/2245500}.
\newblock Publisher: JSTOR.

\bibitem[Finkelstein et~al.(2012)Finkelstein, Taubman, Wright, Bernstein, Gruber, Newhouse, Allen, Baicker, and {Oregon Health Study Group}]{finkelstein_oregon_2012}
Amy Finkelstein, Sarah Taubman, Bill Wright, Mira Bernstein, Jonathan Gruber, Joseph~P. Newhouse, Heidi Allen, Katherine Baicker, and {Oregon Health Study Group}.
\newblock The {Oregon} {Health} {Insurance} {Experiment}: {Evidence} from the {First} {Year}*.
\newblock \emph{The Quarterly Journal of Economics}, 127\penalty0 (3):\penalty0 1057--1106, August 2012.
\newblock ISSN 0033-5533.
\newblock \doi{10.1093/qje/qjs020}.
\newblock URL \url{https://doi.org/10.1093/qje/qjs020}.

\bibitem[Guo and Richardson(2021)]{guo_chernoff-type_2021}
F.~Richard Guo and Thomas~S. Richardson.
\newblock Chernoff-{Type} {Concentration} of {Empirical} {Probabilities} in {Relative} {Entropy}.
\newblock \emph{IEEE Transactions on Information Theory}, 67\penalty0 (1):\penalty0 549--558, January 2021.
\newblock ISSN 1557-9654.
\newblock \doi{10.1109/TIT.2020.3034539}.
\newblock URL \url{https://ieeexplore.ieee.org/abstract/document/9241858}.

\bibitem[Gustafson(2005)]{gustafson_model_2005}
Paul Gustafson.
\newblock On {Model} {Expansion}, {Model} {Contraction}, {Identifiability} and {Prior} {Information}: {Two} {Illustrative} {Scenarios} {Involving} {Mismeasured} {Variables}.
\newblock \emph{Statistical Science}, 20\penalty0 (2):\penalty0 111--140, May 2005.
\newblock ISSN 0883-4237, 2168-8745.
\newblock \doi{10.1214/088342305000000098}.
\newblock URL \url{https://projecteuclid.org/journals/statistical-science/volume-20/issue-2/On-Model-Expansion-Model-Contraction-Identifiability-and-Prior-Information/10.1214/088342305000000098.full}.
\newblock Publisher: Institute of Mathematical Statistics.

\bibitem[Heckman and Urzúa(2010)]{heckman_comparing_2010}
James~J. Heckman and Sergio Urzúa.
\newblock Comparing {IV} with structural models: {What} simple {IV} can and cannot identify.
\newblock \emph{Journal of Econometrics}, 156\penalty0 (1):\penalty0 27--37, May 2010.
\newblock ISSN 0304-4076.
\newblock \doi{10.1016/j.jeconom.2009.09.006}.
\newblock URL \url{https://www.sciencedirect.com/science/article/pii/S0304407609001973}.

\bibitem[Hernán and Robins(2019)]{hernan_causal_2019}
M.~A. Hernán and J.~M. Robins.
\newblock \emph{Causal inference}.
\newblock CRC Boca Raton, FL, 2019.

\bibitem[Imbens(2010)]{imbens_better_2010-1}
Guido~W. Imbens.
\newblock Better {LATE} {Than} {Nothing}: {Some} {Comments} on {Deaton} (2009) and {Heckman} and {Urzua} (2009).
\newblock \emph{Journal of Economic Literature}, 48\penalty0 (2):\penalty0 399--423, June 2010.
\newblock ISSN 0022-0515.
\newblock \doi{10.1257/jel.48.2.399}.
\newblock URL \url{https://www.aeaweb.org/articles?id=10.1257/jel.48.2.399}.

\bibitem[Imbens(2022)]{imbens_causality_2022}
Guido~W. Imbens.
\newblock Causality in {Econometrics}: {Choice} vs {Chance}.
\newblock \emph{Econometrica}, 90\penalty0 (6):\penalty0 2541--2566, 2022.
\newblock ISSN 1468-0262.
\newblock \doi{10.3982/ECTA21204}.
\newblock URL \url{https://onlinelibrary.wiley.com/doi/abs/10.3982/ECTA21204}.
\newblock \_eprint: https://onlinelibrary.wiley.com/doi/pdf/10.3982/ECTA21204.

\bibitem[Ioannidis et~al.(2019)Ioannidis, Tan, and Blum]{ioannidis_limitations_2019}
John~P.A. Ioannidis, Yuan~Jin Tan, and Manuel~R. Blum.
\newblock Limitations and {Misinterpretations} of {E}-{Values} for {Sensitivity} {Analyses} of {Observational} {Studies}.
\newblock \emph{Annals of Internal Medicine}, 170\penalty0 (2):\penalty0 108--111, January 2019.
\newblock ISSN 0003-4819.
\newblock \doi{10.7326/M18-2159}.
\newblock URL \url{https://www.acpjournals.org/doi/full/10.7326/M18-2159}.
\newblock Publisher: American College of Physicians.

\bibitem[Kitagawa(2015)]{kitagawa_test_2015}
Toru Kitagawa.
\newblock A {Test} for {Instrument} {Validity}.
\newblock \emph{Econometrica}, 83\penalty0 (5):\penalty0 2043--2063, September 2015.
\newblock ISSN 1468-0262.
\newblock \doi{10.3982/ECTA11974}.
\newblock URL \url{https://onlinelibrary.wiley.com/doi/10.3982/ECTA11974}.
\newblock Publisher: John Wiley \& Sons, Ltd.

\bibitem[Kowalski(2023)]{kowalski_how_2023}
Amanda~E. Kowalski.
\newblock How to examine external validity within an experiment.
\newblock \emph{Journal of Economics \& Management Strategy}, 32\penalty0 (3):\penalty0 491--509, 2023.
\newblock ISSN 1530-9134.
\newblock \doi{10.1111/jems.12468}.
\newblock URL \url{https://onlinelibrary.wiley.com/doi/abs/10.1111/jems.12468}.

\bibitem[Locher et~al.(2025)Locher, Stensrud, and Sarvet]{locher_interpretational_2025}
Luca Locher, Mats~J. Stensrud, and Aaron~L. Sarvet.
\newblock Interpretational errors with instrumental variables, September 2025.
\newblock URL \url{http://arxiv.org/abs/2509.02045}.
\newblock arXiv:2509.02045 [econ].

\bibitem[Manski(2020)]{manski_lure_2020}
Charles~F. Manski.
\newblock The lure of incredible certitude.
\newblock \emph{Economics \& Philosophy}, 36\penalty0 (2):\penalty0 216--245, July 2020.
\newblock ISSN 0266-2671, 1474-0028.
\newblock \doi{10.1017/S0266267119000105}.
\newblock URL \url{https://www.cambridge.org/core/journals/economics-and-philosophy/article/abs/lure-of-incredible-certitude/A1F09783377B05F20B83543CD40C7639}.

\bibitem[Marbach and Hangartner(2020)]{marbach_profiling_2020}
Moritz Marbach and Dominik Hangartner.
\newblock Profiling {Compliers} and {Noncompliers} for {Instrumental}-{Variable} {Analysis}.
\newblock \emph{Political Analysis}, 28\penalty0 (3):\penalty0 435--444, July 2020.
\newblock ISSN 1047-1987, 1476-4989.
\newblock \doi{10.1017/pan.2019.48}.
\newblock URL \url{https://www.cambridge.org/core/journals/political-analysis/article/profiling-compliers-and-noncompliers-for-instrumentalvariable-analysis/0E2AA51F9BD436A5DB5A84462732A9C3}.

\bibitem[Richardson and Robins(2013)]{richardson_single_2013}
T.~S. Richardson and J.~M. Robins.
\newblock Single world intervention graphs ({SWIGs}): {A} unification of the counterfactual and graphical approaches to causality.
\newblock \emph{Center for the Statistics and the Social Sciences, University of Washington Series. Working Paper}, 128:\penalty0 2013, 2013.

\bibitem[Richardson et~al.(2011)Richardson, Evans, and Robins]{richardson_transparent_2011}
Thomas~S. Richardson, Robin~J. Evans, and James~M. Robins.
\newblock Transparent parameterizations of models for potential outcomes.
\newblock \emph{Bayesian statistics}, 9:\penalty0 569--610, 2011.
\newblock URL \url{https://www.uv.es/bernardo/RichardsonT.pdf}.
\newblock Publisher: Oxford University Press New York.

\bibitem[Sjölander and Greenland(2022)]{sjolander_are_2022}
Arvid Sjölander and Sander Greenland.
\newblock Are {E}-values too optimistic or too pessimistic? {Both} and neither!
\newblock \emph{International Journal of Epidemiology}, 51\penalty0 (2):\penalty0 355--363, April 2022.
\newblock ISSN 0300-5771.
\newblock \doi{10.1093/ije/dyac018}.
\newblock URL \url{https://doi.org/10.1093/ije/dyac018}.

\bibitem[Swanson and Hernán(2013)]{swanson_commentary_2013}
Sonja~A. Swanson and Miguel~A. Hernán.
\newblock Commentary: {How} to {Report} {Instrumental} {Variable} {Analyses} ({Suggestions} {Welcome}).
\newblock \emph{Epidemiology}, 24\penalty0 (3):\penalty0 370, May 2013.
\newblock ISSN 1044-3983.
\newblock \doi{10.1097/EDE.0b013e31828d0590}.
\newblock URL \url{https://journals.lww.com/epidem/fulltext/2013/05000/Commentary___How_to_Report_Instrumental_Variable.00007.aspx}.

\bibitem[Swanson and Hernán(2018)]{swanson_challenging_2018}
Sonja~A Swanson and Miguel~A Hernán.
\newblock The challenging interpretation of instrumental variable estimates under monotonicity.
\newblock \emph{International Journal of Epidemiology}, 47\penalty0 (4):\penalty0 1289--1297, August 2018.
\newblock ISSN 0300-5771.
\newblock \doi{10.1093/ije/dyx038}.
\newblock URL \url{https://doi.org/10.1093/ije/dyx038}.

\bibitem[Swanson et~al.(2018)Swanson, Hernán, Miller, Robins, and Richardson]{swanson_partial_2018}
Sonja~A Swanson, Miguel~A Hernán, Matthew Miller, James~M Robins, and Thomas~S Richardson.
\newblock Partial identification of the average treatment effect using instrumental variables: review of methods for binary instruments, treatments, and outcomes.
\newblock \emph{Journal of the American Statistical Association}, 113\penalty0 (522):\penalty0 933--947, 2018.
\newblock Publisher: Taylor \& Francis.

\bibitem[VanderWeele and Ding(2017)]{vanderweele_sensitivity_2017}
Tyler~J. VanderWeele and Peng Ding.
\newblock Sensitivity {Analysis} in {Observational} {Research}: {Introducing} the {E}-{Value}.
\newblock \emph{Annals of Internal Medicine}, 167\penalty0 (4):\penalty0 268--274, August 2017.
\newblock ISSN 0003-4819, 1539-3704.
\newblock \doi{10.7326/M16-2607}.
\newblock URL \url{https://www.acpjournals.org/doi/10.7326/M16-2607}.

\bibitem[Wang et~al.(2017)Wang, Robins, and Richardson]{wang_falsification_2017}
Linbo Wang, James~M. Robins, and Thomas~S. Richardson.
\newblock On falsification of the binary instrumental variable model.
\newblock \emph{Biometrika}, 104\penalty0 (1):\penalty0 229--236, March 2017.
\newblock ISSN 0006-3444.
\newblock \doi{10.1093/biomet/asw064}.
\newblock URL \url{https://doi.org/10.1093/biomet/asw064}.

\end{thebibliography}

\newpage
\appendix
\section{Identification of type shares and potential outcome means}
\label{app:identification}

Let $A_i \in \{0,1\}$ denote the binary treatment, $Z_i \in \{0,1\}$ the binary instrument, and $Y_i \in \mathbb{R}$ the outcome. Under Assumptions~\eqref{ass:1}--\eqref{ass:3}, the type shares and conditional potential outcome means are identified from the observed data as follows.

\textbf{Type shares.} The share of always-takers is identified from the treatment rate in the control arm, the share of never-takers from the untreated rate in the treatment arm, and the share of compliers as the first-stage effect of the instrument:
\begin{align*}
\pi_A &= P(A_i = 1 \mid Z_i = 0), \\
\pi_N &= P(A_i = 0 \mid Z_i = 1), \\
\pi_C &= P(A_i = 1 \mid Z_i = 1) - P(A_i = 1 \mid Z_i = 0).
\end{align*}

\textbf{Always-takers.} Always-takers have $A_i = 1$ regardless of $Z_i$. Under the exclusion restriction, their observed outcome under $Z_i = 0$ identifies their potential outcome under treatment:
\[
\mu_A^{(1)} = \mathbb{E}[Y_i \mid A_i = 1, Z_i = 0].
\]

\textbf{Never-takers.} Never-takers have $A_i = 0$ regardless of $Z_i$. Under the exclusion restriction, their observed outcome under $Z_i = 1$ identifies their potential outcome under no treatment:
\[
\mu_N^{(0)} = \mathbb{E}[Y_i \mid A_i = 0, Z_i = 1].
\]

\textbf{Compliers.} The complier means are identified by subtracting the always-taker and never-taker contributions from the instrument-stratified outcome means. In the treated arm ($Z_i = 1$), the observed treated outcomes are a mixture of compliers and always-takers:
\[
\mathbb{E}[Y_i \mid A_i = 1, Z_i = 1] = \frac{\pi_C \mu_C^{(1)} + \pi_A \mu_A^{(1)}}{\pi_C + \pi_A},
\]
so that:
\[
\mu_C^{(1)} = \frac{(\pi_C + \pi_A)\,\mathbb{E}[Y_i \mid A_i = 1, Z_i = 1] - \pi_A \mu_A^{(1)}}{\pi_C}.
\]
Similarly, in the control arm ($Z_i = 0$), the observed untreated outcomes are a mixture of compliers and never-takers:
\[
\mathbb{E}[Y_i \mid A_i = 0, Z_i = 0] = \frac{\pi_C \mu_C^{(0)} + \pi_N \mu_N^{(0)}}{\pi_C + \pi_N},
\]
so that:
\[
\mu_C^{(0)} = \frac{(\pi_C + \pi_N)\,\mathbb{E}[Y_i \mid A_i = 0, Z_i = 0] - \pi_N \mu_N^{(0)}}{\pi_C}.
\]
The LATE is then identified as:
\[
\LATE = \mu_C^{(1)} - \mu_C^{(0)}.
\]

All quantities identified above are smooth functions of conditional means of the observed data $(Z_i, A_i, Y_i)$, provided that $\pi_C > 0$, which is guaranteed by the relevance assumption~\eqref{ass:3}. These conditional means can be consistently estimated by their sample analogues, so the identified quantities admit consistent plug-in estimators by the continuous mapping theorem.

\section{Additional notes on the interpretation of thresholds and boundaries}
\label{app:interpretation}
\setcounter{figure}{0}

This appendix illustrates how the threshold and the boundary in the examples of Section~\ref{sec:examples} relate to the identified quantities. Grey bands show pointwise 95\% bootstrap confidence intervals.

\begin{figure}[h!]
\centering
\includegraphics[width=\textwidth]{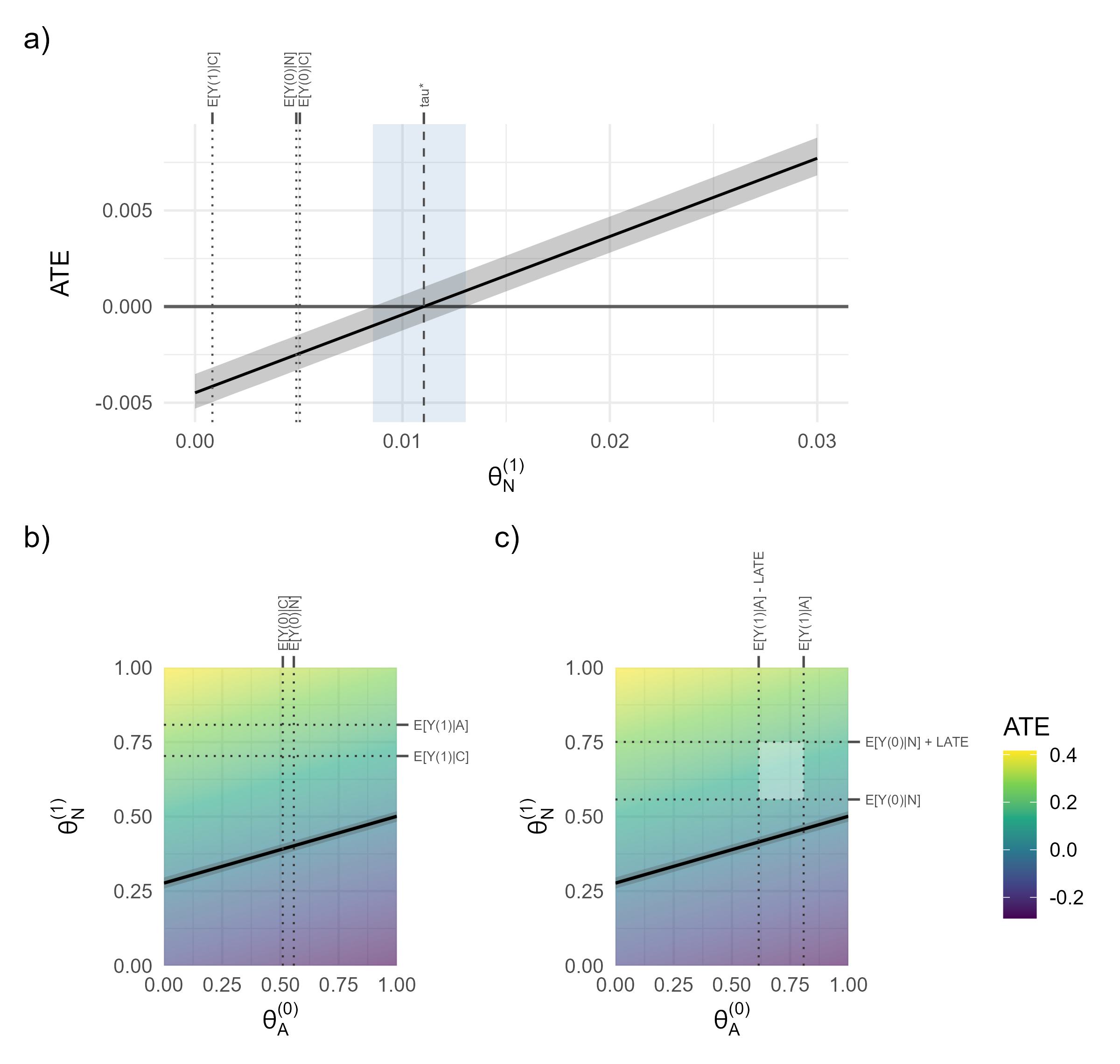}
\caption{Sensitivity analyses for the examples of Section~\ref{sec:examples}, with pointwise 95\% bootstrap confidence bands. (a) Estimated ATE as a function of $\sensNT$ in the example of \cite{bakker_supplemental_2019}; the dashed line and the shaded area show the estimated threshold $\thresholdNTstarhat$ and its 95\% confidence interval, and dotted lines mark the identified potential outcome means. (b, c) Estimated ATE as a function of $\sensAT$ and $\sensNT$ in the example of \cite{finkelstein_oregon_2012}, with the estimated boundary $\hat{\mathcal{B}}$ . Dotted lines mark reference values for (b) outcome heterogeneity and (c) effect heterogeneity between compliance types.}
\label{fig:app_sensitivity}
\end{figure}

\textbf{One-sided non-compliance.} \appfigname~\ref{fig:app_sensitivity}(a) shows the estimated ATE as a function of $\sensNT$ in the example of \cite{bakker_supplemental_2019}. Under one-sided non-compliance, the ATE is linear in $\sensNT$ with slope $\pi_N$ and crosses zero at the threshold $\thresholdNTstarhat$ (dashed line, with its 95\% confidence interval shaded). The decision criterion $\ATE < 0$ holds for all values of $\sensNT$ to the left of the threshold. The dotted lines mark the identified potential outcome means, so that the position of the threshold relative to these lines shows directly which comparisons it implies. Here, the threshold lies to the right of all identified means: the criterion fails only if the cancer incidence of never-takers under MRI screening exceeds all identified potential outcome means, including the never-takers' own incidence without screening.

\textbf{Two-sided non-compliance.}

\appfigname~\ref{fig:app_sensitivity}(b) and (c) show the estimated ATE as a function of both sensitivity parameters in the example of \cite{finkelstein_oregon_2012}. The decision criterion $\ATE > 0$ holds above the estimated boundary $\hat{\mathcal{B}}$ (solid line). The two panels add reference lines that support different arguments.

In panel (b), horizontal lines mark the identified means of the other compliance types under treatment, $\mu_C^{(1)}$ and $\mu_A^{(1)}$, with which $\sensNT$ can be compared, and vertical lines mark the identified means under no treatment, $\mu_C^{(0)}$ and $\mu_N^{(0)}$, with which $\sensAT$ can be compared. These lines show how much outcome heterogeneity between compliance types is required for the criterion to fail. Here, $\hat{\mathcal{B}}$ lies well below both horizontal lines over the entire range of $\sensAT$: the criterion fails only if never-takers have a substantially lower outcome under treatment than compliers and always-takers, although never-takers and compliers have similar outcomes under no treatment.

In panel (c), horizontal lines mark the values of $\sensNT$ at which the effect among never-takers is zero ($\sensNT = \mu_N^{(0)}$) or equal to the LATE ($\sensNT = \mu_N^{(0)} + \LATE$), and vertical lines mark the analogous values of $\sensAT$ for always-takers ($\sensAT = \mu_A^{(1)}$ and $\sensAT = \mu_A^{(1)} - \LATE$). These lines show how much effect heterogeneity between compliance types is required for the criterion to fail. The highlighted rectangle contains all values for which the effects among always-takers and never-takers lie between zero and the LATE; it lies entirely above $\hat{\mathcal{B}}$. Moreover, $\hat{\mathcal{B}}$ lies below the line $\sensNT = \mu_N^{(0)}$ over the entire range of $\sensAT$: the criterion holds whenever the treatment does not decrease the outcome among never-takers, irrespective of the effect among always-takers.

\section{Equivalence with Balke--Pearl bounds}
\label{app:bp}

\textbf{Notation.} Following \cite{balke_bounds_1997}, for binary outcomes $Y_i \in \{0,1\}$ let
\[
p_{ya.z} = P(Y_i = y, A_i = a \mid Z_i = z), \quad y, a, z \in \{0,1\}.
\]
The type shares and identified potential outcome means introduced in \ref{app:identification} can be expressed in terms of these observable quantities as:
\begin{align*}
\pi_A &= p_{11.0} + p_{01.0}, \qquad \pi_N = p_{00.1} + p_{10.1}, \qquad \pi_C = (p_{11.1} + p_{01.1}) - (p_{11.0} + p_{01.0}), \\
\mu_A^{(1)} &= \frac{p_{11.0}}{\pi_A}, \qquad \mu_N^{(0)} = \frac{p_{10.1}}{\pi_N}, \qquad \mu_C^{(1)} = \frac{p_{11.1} - p_{11.0}}{\pi_C}, \qquad \mu_C^{(0)} = \frac{p_{10.0} - p_{10.1}}{\pi_C}.
\end{align*}

\cite{balke_bounds_1997} write that under Assumptions~\eqref{ass:1}--\eqref{ass:3},
\[
\max\left\{\begin{array}{l}
p_{00.0} + p_{11.1} - 1 \\
p_{00.1} + p_{11.1} - 1 \\
p_{11.0} + p_{00.1} - 1 \\
p_{00.0} + p_{11.0} - 1
\end{array}\right\}
\leq \ATE \leq
\min\left\{\begin{array}{l}
1 - p_{10.0} - p_{01.1} \\
1 - p_{01.0} - p_{10.1} \\
1 - p_{01.0} - p_{10.0} \\
1 - p_{01.1} - p_{10.1}
\end{array}\right\}
\]

\textbf{Claim.} Under Assumptions~\eqref{ass:1}--\eqref{ass:5} and binary outcomes $Y_i \in \{0,1\}$, the Balke--Pearl sharp bounds on the ATE coincide with the bounds obtained by imputing the most extreme feasible values for $\sensAT$ and $\sensNT$:
\begin{align*}
\max\{p_{00.0} + p_{11.1} - 1,\; p_{00.1} + p_{11.1} - 1,\; p_{11.0} + p_{00.1} - 1,\; p_{00.0} + p_{11.0} - 1\} &= \ATE(1,0), \\
\min\{1 - p_{10.0} - p_{01.1},\; 1 - p_{01.0} - p_{10.1},\; 1 - p_{01.0} - p_{10.0},\; 1 - p_{01.1} - p_{10.1}\} &= \ATE(0,1).
\end{align*}

We first establish four inequalities that follow directly from the identification results and hold throughout. Since $\pi_C \geq 0$ and $\mu_C^{(0)}, \mu_C^{(1)} \in [0,1]$:
\begin{align}
p_{00.0} - p_{00.1} &= \pi_C(1 - \mu_C^{(0)}) \geq 0, \label{ineq:1} \\
p_{11.1} - p_{11.0} &= \pi_C\mu_C^{(1)} \geq 0, \label{ineq:2} \\
p_{01.1} - p_{01.0} &= \pi_C(1 - \mu_C^{(1)}) \geq 0, \label{ineq:3} \\
p_{10.0} - p_{10.1} &= \pi_C\mu_C^{(0)} \geq 0. \label{ineq:4}
\end{align}

\textbf{Equivalence of lower bound.} Substituting $\sensAT = 1$ and $\sensNT = 0$ into equation~\eqref{eq:ate_decomp} and using the identification expressions:
\begin{align*}
\ATE(1,0) &= \pi_C\mu_C^{(1)} - \pi_C\mu_C^{(0)} + \pi_A\mu_A^{(1)} - \pi_A - \pi_N\mu_N^{(0)} \\
&= (p_{11.1} - p_{11.0}) - (p_{10.0} - p_{10.1}) + p_{11.0} - (p_{11.0} + p_{01.0}) - p_{10.1} \\
&= p_{11.1} - p_{11.0} - p_{10.0} - p_{01.0}.
\end{align*}
Since $p_{00.0} + p_{10.0} + p_{01.0} + p_{11.0} = 1$ implies $p_{11.0} + p_{10.0} + p_{01.0} = 1 - p_{00.0}$, substituting gives $\ATE(1,0) = p_{11.1} - (1 - p_{00.0}) = p_{00.0} + p_{11.1} - 1$.

The remaining three lower bound expressions are dominated:
\begin{itemize}
\item $p_{00.1} + p_{11.1} - 1 \leq p_{00.0} + p_{11.1} - 1$ iff $p_{00.1} \leq p_{00.0}$, which holds by~(\ref{ineq:1}).
\item $p_{11.0} + p_{00.1} - 1 \leq p_{00.0} + p_{11.1} - 1$ iff $p_{11.0} - p_{11.1} \leq p_{00.0} - p_{00.1}$, i.e.\ $-\pi_C\mu_C^{(1)} \leq \pi_C(1-\mu_C^{(0)})$, which simplifies to $0 \leq \pi_C(1 - \mu_C^{(0)} + \mu_C^{(1)})$, holding by~(\ref{ineq:1}) and~(\ref{ineq:2}).
\item $p_{00.0} + p_{11.0} - 1 \leq p_{00.0} + p_{11.1} - 1$ iff $p_{11.0} \leq p_{11.1}$, which holds by~(\ref{ineq:2}).
\end{itemize}
Therefore $\max\{p_{00.0} + p_{11.1} - 1,\; p_{00.1} + p_{11.1} - 1,\; p_{11.0} + p_{00.1} - 1,\; p_{00.0} + p_{11.0} - 1\} = \ATE(1,0)$.

\textbf{Equivalence of upper bound.} Substituting $\sensAT = 0$ and $\sensNT = 1$ into equation~\eqref{eq:ate_decomp} and using the identification expressions:
\begin{align*}
\ATE(0,1) &= \pi_C\mu_C^{(1)} - \pi_C\mu_C^{(0)} + \pi_A\mu_A^{(1)} + \pi_N - \pi_N\mu_N^{(0)} \\
&= (p_{11.1} - p_{11.0}) - (p_{10.0} - p_{10.1}) + p_{11.0} + (p_{00.1} + p_{10.1}) - p_{10.1} \\
&= p_{11.1} + p_{00.1} + p_{10.1} - p_{10.0}.
\end{align*}
Since $p_{11.1} + p_{01.1} + p_{10.1} + p_{00.1} = 1$ implies $p_{11.1} + p_{00.1} + p_{10.1} = 1 - p_{01.1}$, substituting gives $\ATE(0,1) = 1 - p_{01.1} - p_{10.0} = 1 - p_{10.0} - p_{01.1}$.

The remaining three upper bound expressions dominate:
\begin{itemize}
\item $1 - p_{01.0} - p_{10.1} \geq 1 - p_{10.0} - p_{01.1}$ iff $p_{10.0} - p_{10.1} \geq p_{01.0} - p_{01.1}$, i.e.\ $\pi_C\mu_C^{(0)} \geq -\pi_C(1-\mu_C^{(1)})$, which simplifies to $0 \leq \pi_C(\mu_C^{(0)} + 1 - \mu_C^{(1)})$, holding by~(\ref{ineq:3}) and~(\ref{ineq:4}).
\item $1 - p_{01.0} - p_{10.0} \geq 1 - p_{10.0} - p_{01.1}$ iff $p_{01.1} \geq p_{01.0}$, which holds by~(\ref{ineq:3}).
\item $1 - p_{01.1} - p_{10.1} \geq 1 - p_{10.0} - p_{01.1}$ iff $p_{10.0} \geq p_{10.1}$, which holds by~(\ref{ineq:4}).
\end{itemize}
Therefore $\min\{1 - p_{10.0} - p_{01.1},\; 1 - p_{01.0} - p_{10.1},\; 1 - p_{01.0} - p_{10.0},\; 1 - p_{01.1} - p_{10.1}\} = \ATE(0,1)$.

\section{Data for the examples}
\label{app:counts}
\setcounter{table}{0}

With a binary instrument, treatment, and outcome, the observed data are fully summarized by the counts of the eight combinations of $(Z_i, A_i, Y_i)$. \apptabname~\ref{tab:counts} reports these counts for both examples in Section~\ref{sec:examples}; all estimates, bounds, thresholds, and bootstrap confidence intervals reported there can be reproduced from them.

In Example 1 \citep{bakker_supplemental_2019}, $Z_i$ indicates assignment to the group invited to supplemental MRI screening, $A_i$ indicates participation in MRI screening, and $Y_i$ indicates a diagnosis of interval breast cancer. In Example 2 \citep{finkelstein_oregon_2012}, $Z_i$ indicates selection in the lottery, $A_i$ indicates enrollment in Medicaid, and $Y_i$ indicates any primary care visit in the previous 12 months, as reported by respondents to the 12-month survey.

\begin{table}[t]
\centering
\begin{tabular}{@{}cccrr@{}}
\toprule
$Z_i$ & $A_i$ & $Y_i$ & Example 1 & Example 2 \\
\midrule
0 & 0 & 0 & 32,151 & 4,717 \\
0 & 0 & 1 & 161 & 5,572 \\
0 & 1 & 0 & 0 & 292 \\
0 & 1 & 1 & 0 & 1,230 \\
1 & 0 & 0 & 3,262 & 2,971 \\
1 & 0 & 1 & 16 & 3,745 \\
1 & 1 & 0 & 4,779 & 1,315 \\
1 & 1 & 1 & 4 & 3,650 \\
\midrule
\multicolumn{3}{@{}l}{Total} & 40,373 & 23,492 \\
\bottomrule
\end{tabular}
\caption{Number of individuals by instrument $Z_i$, treatment $A_i$, and outcome $Y_i$ in the examples of \cite{bakker_supplemental_2019} (Example 1) and \cite{finkelstein_oregon_2012} (Example 2).}
\label{tab:counts}
\end{table}

\end{document}